\documentstyle[psfig,epsfig]{aipproc}

\def\etal{{\it et al.}}

\begin{document}

%\begin{flushright}
%hep-ph/0207234 \\
%UCCHEP/15-00
%\end{flushright}
%\vskip -.7cm

\title{Neutrinos in Supersymmetry without R-Parity\footnote{
Talk given by M.A.D. at High Energy Physics Workshop, Valdivia, 
Chile.}}

\author{M. Hirsch$^{a}$, M.A. D\'\i az$^{b}$, W. Porod$^{c}$, 
J.C. Rom\~ao$^{d}$, J.W.F. Valle$^{a}$}
\address{
$^a$Astroparticle and HEP Group, IFIC, Ed. Inst. de Invest.,
c.p. 22085, E46071 Valencia, Spain\\
$^b$Departamento de F\'\i sica, Universidad Cat\'olica de Chile,
Macul, Santiago 6904411, Chile\\
$^c$Inst. f\"ur Theor. Physik, Universit\"at Z\"urich, Switzerland\\
$^d$Dept. de F\'\i sica, Instituto Superior T\'ecnico,
Av. Rovisco Pais 1, 1049-001 Lisboa, Portugal
}

%\lefthead{LEFT head}
%\righthead{RIGHT head}
\maketitle

\vskip -.6cm
\begin{abstract}

We show how Bilinear R-Parity violation within the Minimal Supersymmetric 
Standard Model can solve the atmospheric and solar neutrino problems by 
generating naturally small and hierarchical neutrino masses, together 
with neutrino mixing angles consistent with experiments. The relation
between collider and neutrino physics is emphasized.

\end{abstract}

It is well understood that the neutrino sector is a window to new physics 
beyond the Standard Model (SM). Several experimental results indicate
that neutrinos oscillate and have mass\cite{review}. Here we report
on a supersymmetric solution to the atmospheric and solar neutrino 
anomalies in terms of a hierarchical neutrino mass spectrum\cite{us,more}. 
A deficit of muon atmospheric neutrinos \cite{atm} can be explained by 
oscillations between muon and tau neutrinos with a mass squared difference
\begin{equation}
\Delta m^2_{atm}=m_3^2-m_2^2\approx 10^{-3}-10^{-2}\,{\mathrm{eV}}^2\,,
\end{equation}
and with near maximal mixing. Solar neutrinos detected in underground 
experiments are less than expected from theory, and this anomaly can be 
explained by oscillations between electron and muon neutrinos with a 
large mixing angle, solution favored specially by SNO results\cite{SNO},
and a squared mass difference
\begin{equation}
\Delta m^2_{sol}=m_2^2-m_1^2\approx 10^{-5}-10^{-4}\,{\mathrm{eV}}^2\,.
\end{equation}
In supersymmetry, one possible way to generate neutrino masses is via 
bilinear R-Parity violation (BRpV)\cite{DRV}, where bilinear terms, which
violate R-Parity and lepton number, are added to the MSSM superpotential:
\begin{equation}
W=W_{MSSM}+\epsilon_i\hat L_i\hat H_2
\end{equation}
Their presence is motivated by models where R-Parity and lepton
number are spontaneously broken by right handed neutrino
vev's\cite{spon}, or by models with horizontal symmetries\cite{hori}.
These bilinear terms\footnote{
Trilinear R-Parity violation, if present, also contribute to neutrino 
masses\cite{trilinears}
} induce sneutrino vacuum expectation values $v_i$ and mixing 
between neutrinos and neutralinos. With a low energy see-saw mechanism, 
this mixing generates at tree level a mass to one neutrino, while the other 
two remain massless. This tree level neutrino mass satisfy 
$m_3\sim|\vec\Lambda|^2$, where $\Lambda_i=\mu v_i+\epsilon_i v_1$, which 
are naturally small parameters in models with universal soft parameters at 
the GUT scale:
\begin{equation}
\Lambda_i\sim{{3h_b^2|\vec\epsilon|}\over{16\pi^2}}m_Z
\ln{{M_{GUT}^2}\over{m_Z^2}}
\end{equation}
Masses for the other two neutrinos are generated once we include the 
one-loop corrections to the neutralino/neutrino mass matrix.

\begin{figure}
\setlength{\unitlength}{1mm}
\begin{picture}(50,70)
\put(-42,-150)
{\mbox{\epsfig{figure=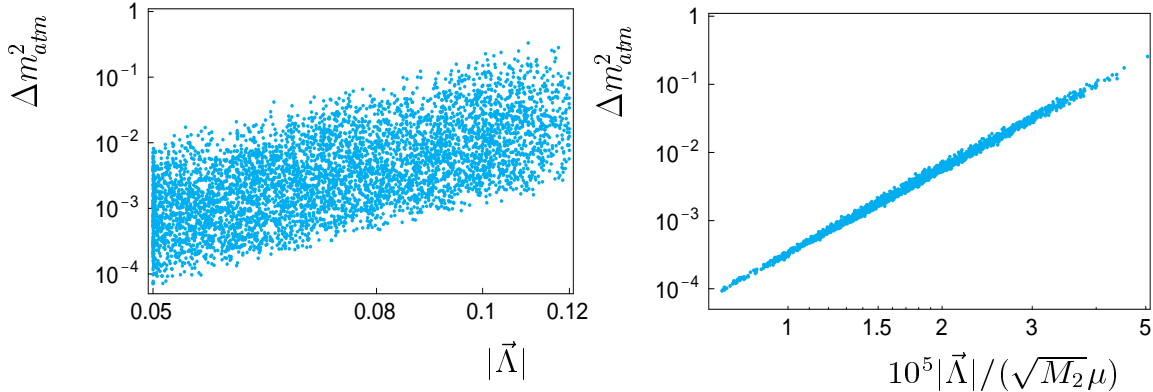,height=28.0cm,width=24.0cm}}}
\end{picture}
\caption[]{\it Example of calculated $\Delta m^2_{atm}$ as a function of 
  (left) the alignment parameter ${\vec \Lambda}$ and (right), as
  function of $|{\vec \Lambda}|/(\sqrt{M_2} \mu)$. }
\label{atmmass}
\end{figure}
In Fig.~\ref{atmmass} we see the strong correlation between the $\Lambda_i$
parameters and the atmospheric neutrino mass, which is generated by the tree 
level mass matrix:
\begin{equation}
\sqrt{\Delta m_{atm}^2}\approx m_3\sim |\vec\Lambda|^2/M_{SUSY}
\end{equation}
\begin{figure}
\setlength{\unitlength}{1mm}
\begin{picture}(50,70)
\put(-42,-150)
{\mbox{\epsfig{figure=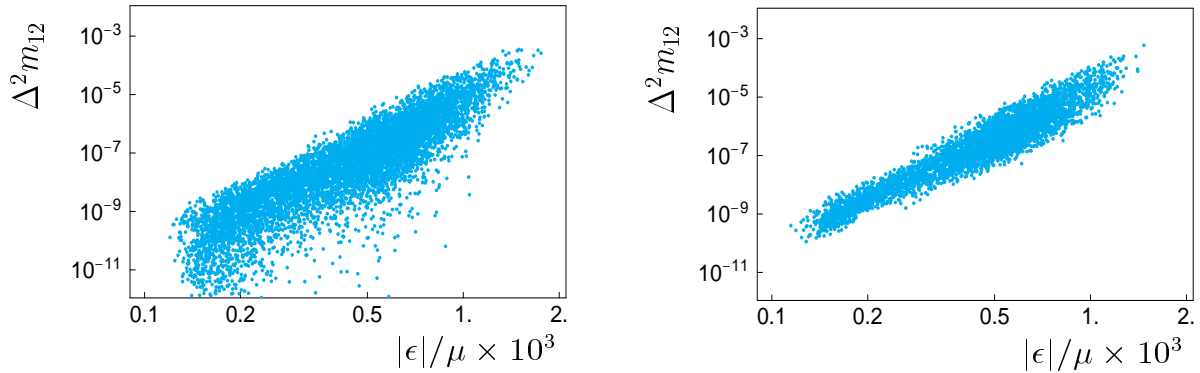,height=28.0cm,width=24.0cm}}}
\end{picture}
\caption[]{\it Solar mass squared difference as a function of 
$|\epsilon|/\mu$ for $\mu \le 0$ (left)  and $\mu > 0$ (right).}
\label{delm12vmu}
\end{figure}
On the other hand, in Fig.~\ref{delm12vmu} we see the correlation between 
the solar mass and the $\epsilon_i$ parameters, which appear in the 
neutralino/neutrino mass matrix at one loop via couplings, showing that
the solar mass is a genuinely one-loop effect. The main contribution comes 
from sbottom loops, and we have approximately:
\begin{equation}
\sqrt{\Delta m_{sol}^2}\approx m_2\sim{{3h_b^2|\vec\epsilon|^2}
\over{16\pi^2}}{{m_b}\over{M_{SUSY}^2}}\ln{{M_{\tilde b_2}^2}\over
{M_{\tilde b_1}^2}}\,.
\end{equation}
Maximal atmospheric and solar mixing angles can be obtained in this model
relaxing exact universality. Maximal atmospheric angle is obtained when
\begin{equation}
\tan\theta_{23}=\Lambda_{\mu}/\Lambda_{\tau}\approx 1
\end{equation}
and the Chooz constraint $\sin^2\theta_{13}<0.045$\cite{Chooz} can be 
satisfied taking
\begin{equation}
|\tan\theta_{13}|=|\Lambda_e|/\sqrt{\Lambda_{\tau}^2+\Lambda_{\mu}^2}
\ll 1\,.
\end{equation}
Finally, after relaxing exact universality of soft mass parameters the
maximal mixing can be achieved, for example, taking 
\begin{equation}
\epsilon_e\sim\epsilon_{\mu}\,.
\end{equation}

\begin{figure}
\setlength{\unitlength}{1mm}
\begin{picture}(50,90)
\put(40,30)
{\mbox{\epsfig{figure=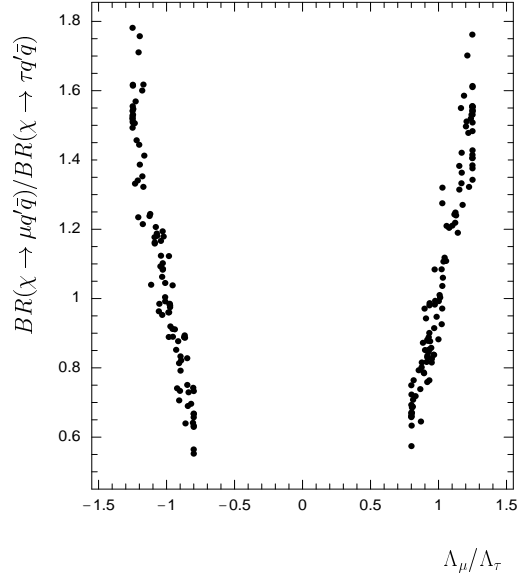,height=8.0cm}}}
\end{picture}
\caption[]{\it Ratio of branching ratios for semileptonic LSP
decays into muons and taus: $BR(\chi \to \mu q' \bar q)/ BR(\chi \to
\tau q' \bar q$) as function of $\Lambda_\mu/\Lambda_\tau$.}
\label{lettfig4}
\end{figure}

A very interesting property of the BRpV model is that neutrino physics is 
closely related to collider physics in a measurable way. As mentioned
before, mixing between 
neutrinos and neutralinos allow the generation of neutrino masses through 
a low energy see-saw mechanism. The atmospheric mass depends mainly on the
$\Lambda_i$ parameters, and the same parameters control the mixing between
gauginos and neutrinos. In this way, the neutralinos will have decay 
channels into leptons with couplings that depend on the $\Lambda_i$. In 
the case of the lightest neutralino, lepton number and R-parity violating 
decays branching ratios sum up to 100\%, because R-Parity conserving decays 
are closed. Since neutralino couplings to leptons will depend on the 
$\Lambda_i$, measurements on branching ratios will give information on 
neutrino physics and vice versa. In Fig.~\ref{lettfig4} we plot a ratio of 
semileptonic branching ratios of the lightest neutralino decays into muons 
and taus. It is evident from the figure the correlation between the former
ratio and $\Lambda_{\mu}/\Lambda_{\tau}$, quantity that controls the 
atmospheric mass scale.

In summary, lepton number and R-Parity violating bilinear terms in 
supersymmetric models can generate hierarchical neutrino masses and bimaximal
mixings, and thus explain the solar and atmospheric neutrino anomalies.
This model is highly predictive, introducing very few extra parameters
to the MSSM, while predicting three masses and three mixing angles. This
model has the additional interesting feature of connecting neutrino physics 
with collider physics. For example, measurements on branching ratios of
neutralino decays give information on neutrino parameters and vice versa.

%\newpage
%

\end{document}